# Carbon nanotube/metal corrosion issues for nanotube coatings and inclusions in a matrix

**Farhad Daneshvar**

Polymer Technology Centre, Department of Materials Science and Engineering, Texas A&M University, College Station, TX 77843, USA. E-mail: f.daneshvar@tamu.edu

March 2016

# Abstract

Corrosion is an inevitable phenomenon that is inherent in metals and even though there has been significant research on this subject, no ideal protection has been discovered to fully prevent corrosion. However, methods such as using protective coatings, and modifying the structure or composition of the material have been used to slow down gradual corrosion and fortunately they proved to be quite beneficial. The research focus has shifted to integrating novel materials and structures to improve the corrosion resistance of composites. Carbon nanotubes (CNTs) are an attractive and promising filler due to their chemical inertness and high mechanical, electrical, and thermal properties. CNTs can fill the gaps of metals and polymer-based composites by forming a passive layer on metals and promoting sacrificial protection in zinc rich polymer (ZRP) coatings, and can therefore function as an anti-corrosion filler. This paper reviews the research that has been performed to better understand the influence of CNTs on corrosion resistance in composites. Accordingly, in metal matrix composites (MMCs), most of the work has been carried out on electrodeposited coatings, namely Ni-based-CNT composites, which show improved corrosion resistance by CNT addition. On the other hand, there are a few papers that have studied the corrosion resistance of Mg-based-CNT composites and their corrosion results contradict those obtained from other metal-CNT composites. For ZRPs or polymer-based coatings there are a few papers that studied the effect of CNTs on the corrosion of said composites. It is believed that CNTs can strengthen the adhesion between the coating and the substrate and facilitate sacrificial protection by Zn particles by forming a conductive network, hence the improved corrosion resistance. It should be noted that this research topic is in its infancy stage and the prevalence of papers that have studied the mechanism is lacking.

# Contents

# Table of Figures

## 1. Introduction

Since the introduction of carbon nanotubes (CNTs) by Iijima in 1991 [1], they become a topic of many research. This stems from carbon nanotube's remarkable properties such as optical properties, extremely high mechanical properties, high thermal and electrical conductivity and its unique structure which at first attracted the attention of researchers. Many potential applications were mentioned for CNTs, from energy storage [2] to electronics [3] and spintronics [4] however one particular application is using CNTs as the reinforcement material in composites inspired by their high Young's modulus (up to 1 TPa) and high tensile strength (about 60 GPa) [5].

Compared with other composites, polymer matrix composites containing CNT have been under more attention [6]. This can be attributed to the remarkable effects of CNTs on mechanical properties and electrical conductivity of polymers. Moreover, there are some challenges in utilizing CNTs in metal matrix composites such as poor wettability of carbon by molten metal, formation of interfacial reaction product and associated galvanic corrosion [7]. It is worth noting that polymer processing is relatively easy and done at lower temperatures compared with metals and ceramics. For metal matrix composites (MMC) in addition to improvement of mechanical properties, CNTs can potentially enhance sensing performance, wear resistance, heat dissipation, and electrochemical performance. Inspired by excellent mechanical and lubricating properties, investigations were carried out on incorporation of CNT in coatings. Facile processing methods such as electrodeposition, electroless deposition, spraying, molecular level mixing and sputtering paved the way of using CNTs in these systems. Although some very excellent results has been reported, improvement in properties of the composites containing CNTs are not as remarkable as expected for most cases. This is mostly attributed to lack of achieving a uniform dispersion of filler within the matrix. Due to high surface energy, as-produced CNTs tend to entangle and form agglomerates, hence do not yield a uniform dispersion which significantly deteriorates the composite properties [8]. To tackle this problem, surface of the CNTs are usually modified by covalent (such as acid treatment) or non-covalent methods (such as surfactants).

Corrosion is an inevitable phenomena and as new composites are developed, their corrosion resistance should be a matter of concern. Sine CNTs have a high chemical stability they can be a promising option to improve the corrosion resistance as a filler in composites [9]. However, unfortunately these research usually study the corrosion properties along with mechanical, wear, electrical and thermal properties of the new composites hence, there are scarce detailed discussions about the effects of CNTs on the corrosion resistance of composites.

Aim of this work is to review the papers that have evaluated the application of CNTs as strengthening phase in composites on corrosion properties and studying the nano/micro effects and interactions between carbon nanotube materials and metal matrices or substrates. The focus will be on protective nanocomposite coatings. Firstly, a brief introduction of corrosion and common methods for evaluating the corrosion resistance of a material is discussed. Subsequently, the effect of CNT incorporation in metallic and polymeric matrices are studied. Finally a few papers that have worked on CNT coatings (that cannot be categorized as polymeric or metallic matrices) are reviewed.

## 2. Corrosion protection and evaluation

Corrosion is a costly and inevitable phenomenon that is inherent in metals and even though there have been significant research on this subject, yet no ideal protection was discovered to prevent it [10]. However, methods such as cathodic/anodic protection, using inhibitors, modifying the material/composition and applying protective coatings have been used for a long time to slow down the gradual corrosion and fortunately they proved to be quiet beneficial.

Corrosion protective coatings can be organic or inorganic or a mixture of both. Using chromates used to be a very efficient method however, due to environmental concerns, they were replaced. Corrosion resistance metals such as Ni and Ti or noble metals such as Au and Ag are the other options. Another type of metallic coating is anodic coatings such as Zn which have lower electrode potential. These coatings are electrochemically more active than the substrate therefore in the presence of a corrosive medium they will oxidize faster and the electrons produced by the coating will flow to the substrate, making it a cathode and thus it will be protected. Organic coatings such as epoxy and polyurethane are also very common. They will hinder the corrosion by making barrier for contact between corrosive agent and the substrate. These polymers are usually mixed with high amount of Zn particles to give the highest performance. CNTs can also be incorporated in these coatings to improve the mechanical properties and wear resistance of these coatings.

Modifying the material can be done by alloying or producing a composite. Owning to their being chemically inert, excellent mechanical morphological and electrical properties, CNTs are an attractive option for corrosion protection applications.

Generally, evaluating the corrosion is done by visual inspections, gravimetric and electrochemical methods. Also for polymer base coatings assessing the strength of bonding of the coating to the substrate, and salt spray test are common. In gravimetric method, the sample is kept in the corrosive environment and periodically its weight is measured. This test usually need to be done in long times. In contrast, electrochemical tests usually take shorter times. In the literature measuring the open circuit potential (OCP) and polarization tests are the most common ones. However, it should be mentioned that although OCP may give an idea of thermodynamic aspect of corrosion, it cannot be used as a criterion for determining the corrosion rate. Electrochemical impedance spectroscopy (EIS) is another common method for evaluating the corrosion of CNT containing composites in the literature. This test can offer details about the mechanism of corrosion, however, especially for polymers in which the corrosive liquid can infiltrate through the coating over time, the test should be done periodically for longer times to give a meaningful result. As it was mentioned, in continuous contact, corrosive liquid can infiltrate through the polymeric coating and reach the substrate. In that case the substrate oxidizes and corrosion products will decrease the bonding of the coating to the substrate. If the coating is easily debonded, failure due to corrosion especially localized corrosion will be very likely therefore, analyzing the bonding strength of polymeric coatings plays a crucial role in determining protection capability of the coating.

### 3. Effect of CNT on corrosion resistance of metal matrix composites

As mentioned before remarkable properties of CNTs lead to a broad interest in using them as strengthening phase into metal matrix composites (MMC). The most common processing methods for making MMCs are electrodeposition, electroless deposition, powder metallurgy, chemical vapor deposition (CVD), melt mixing and molecular level mixing. Although there are some papers that have studied the corrosion resistant of the bulk composites however, most of the research on this field is carried out on MMC-CNT coatings.

Among different MMC-CNT systems, electrodeposited Ni-CNT is one of the most popular ones probably due to its vast application in industry and simplicity of the process. According to literature, integration of CNTs can enhance hardness, corrosion, mechanical and wear properties of nickel base nanocomposite coatings. However, there are scarce reports that has studied corrosion resistance mechanism of Ni-CNT composite coatings thoroughly. Chen et al. [11] used acid treatment and cetyl trimethy ammonium bromide (C-TAB) surfactant to disperse CNTs in water. They reported that integrating these CNTs into Ni coating by electrodeposition improves the corrosion three times. They also observed that the Ni-CNT coating is more resistance to pitting corrosion and has a more stable and positive corrosion potential, $E_{corr}$, compared with Ni coating. CNTs were embedded deeply in Ni and filled the coating's defect and microholes. Also their uniform dispersion promoted the uniform anode polarization and inhibited localized corrosion. This improvement in Ni-CNT coatings corrosion has been reported by other researchers using gravimetric, polarization and EIS tests [8, 11-13]. It is worth noting that in all the works, addition of CNT does not change the shape of EIS curves meaning the corrosion mechanism does not change.

Moreover, it has been shown that CNT incorporation has a positive effect on corrosion resistance of Ni-W/MCNT [14], Ni-Cu-P/MWNT [15] and in a few works Ni-P/CNT [9, 16-18] coatings.

Chen et al. [11] are among the first groups that studied the effect of CNT integration on corrosion resistance of nickel base composite produced by electrodepostion. They used acid treated CNTs modified with a cationic surfactant (cetyl trimethy ammonium bromide (C-TAB)). Weight loss measurements and polarization curves showed that the CNT addition improved the corrosion resistance of nickel coating 3 times. Moreover they observed that CNTs increased pitting corrosion potential and hence improved resistance to pitting corrosion of Ni coating. They attributed this improvement to factors: 1- CNTs fill the holes and gaps of the coating and act as a physical barrier. 2- Uniform dispersion of CNTs (which have higher standard potential) make many corrosion micro cells which will promote Ni (which has lower standard potential) polarization. Shi et al. [18] also observed that carbon nanotubes have a major role on passivation of Ni-P-CNT coating. By addition of CNT, OCP of the Ni-P coating in 0.1 M NaCl solution shifted to more negative values which is rare in the literature. They also reported that incorporation of CNTs in Ni-P coating by electrodeposition can improve the corrosion resistance by promoting formation of passive film. Moreover, EIS curves' shape, in different potentials, did not change with addition of CNTs hence it was concluded that the CNT does not have any effect on mechanism of corrosion. However, to obtain a clear opinion about the effect of CNT on the corrosion mechanism in metals, it is better to do the analysis at OCP and in longer periods.

Recently a more detailed corrosion study was carried out on electroless plated Ni-P-CNT composite. Alishahi et al. [16] noticed that rate of cathodic half reaction is higher for the CNT containing coating. They believed that the cathodic reaction is hydrogen evolution and its rate is increased in the presence of CNTs which can be observed by decrease in cathodic Tafel slope. This not only shifts the corrosion potential to more positive values but also enhances the anodic reaction which is dissolution of Ni. By dissolution of Ni, a phosphorous-rich passivation layer is quickly formed which hinders the corrosion. However, since they used natural water with NaCl, this cathodic reaction is very unlikely to occur and instead the reduction reaction can actually be creation of hydroxyl ions (Eq. 1). Moreover, it has been shown that in in chlorinated water solutions reductions of chlorine species are the major cathodic reactions which is mass transport controlled [19].

$$O_2 + 2\ H_2O + 4\ e^- \rightarrow 4\ OH^- \tag{1}$$

EIS data showed that CNT incorporation increased charge transfer resistance which is attributed to formation of a P-rich passivation layer which blocks the active surface of the coating from the solution and corrosive elements (Fig. 1). Moreover, based on decrease of double layer capacity ($CPE_{dl}$) and increase of $n_{dl}$ to one, they claimed that the CNT can reduce the defects in the coatings by filling the gaps and micro-holes [16].
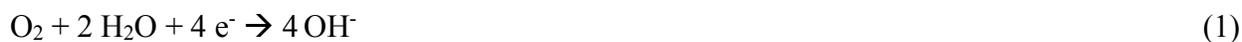

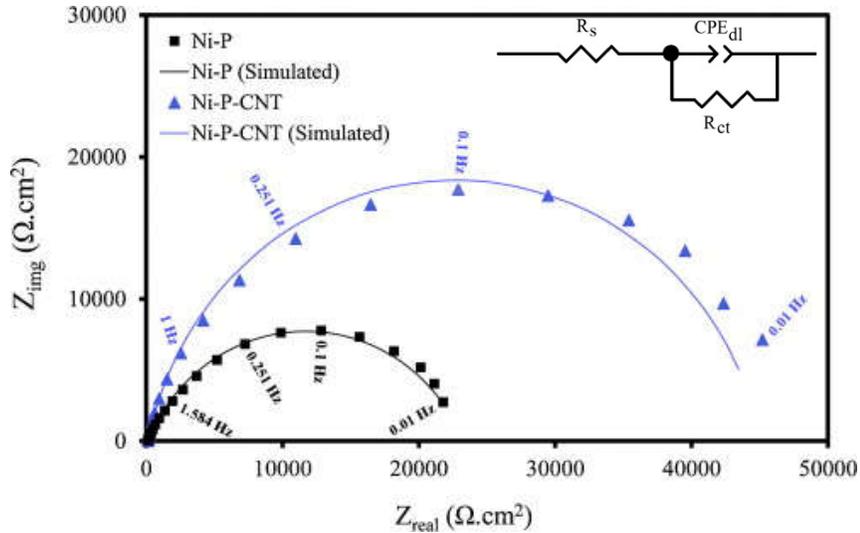

Figure 1.The Nyquist plots of the coatings in 3.5% NaCl solution with the equivalent circuit [16].

Surface modification technique, dispersion, and interfacial adhesion to the matrix play major roles on the corrosion resistance of a composite material. For instance Kim and Oh [20] observed that poor adhesion of CNTs to Ni matrix resulted in a porous structure (Fig. 2) and as the CNT content increases the coating become more porous thus more surface area is exposed to the corroding medium resulting in decrease in corrosion resistance and the potential corrosion shift to more negative values. Guo et al. [5] studied the effect of surfactant on mechanical properties and corrosion resistance of Ni-CNT coatings. While using SDS (anionic) surfactant improved both mechanical properties and corrosion resistance of the coating, application of C-TAB (cationic) resulted in inferior mechanical and electrochemical performance. They attributed this to decreased boundary adhesion of CNTs to Ni matrix.

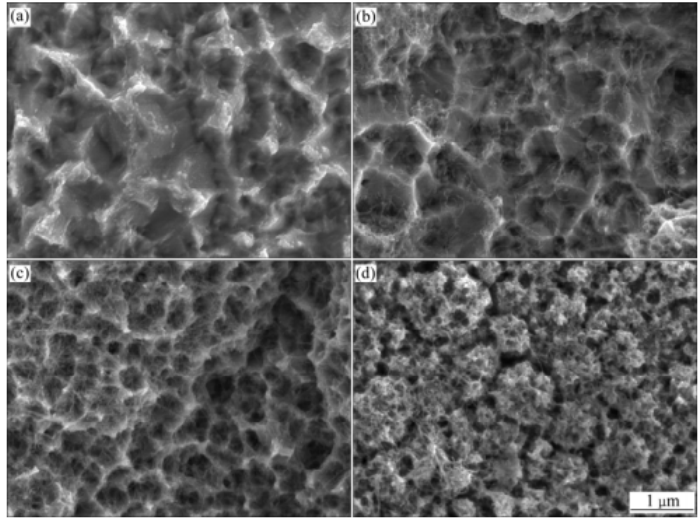

Figure 2. FESEM micrographs of Ni-CNT nanocomposites electrodeposited in solution containing CNT concentration of 1 g/L (a), 2 g/L, (b), 5 g/L (c) and 10 g/L (d) [20].

Also the intensity of CNT acid treatment can affect the corrosion properties of a coating. Harsh acid treatment will introduce many defects on the surface of CNT which will decrease its conductivity significantly. These CNTs are buried within nickel matrix, in contrary with mildly treated CNTs which can reduce CNTs on their surface so will adherently incorporated in the matrix. Therefore, utilizing a harsh acid treatment on CNTs results in inferior mechanical and corrosion properties in metallic composites [13].

Corrosion studies on Cu base-CNT composites are scarce and the result of CNT addition reported to be different. While Zygon [21] observed CNT will deteriorate corrosion resistance, Gill and Monroe [22] reported CNT will decrease the corrosion rate and pitting corrosion. The positive effect was attributed to formation of carbides due to high temperature processing method (PM) which will promote passivation.

For Mg base composites the results are even more controversial. Although at first it was reported that has beneficial effects on corrosion resistance of Mg alloy (AZ91D)-CNT composite since they can keep the oxide layer detached from the substrate thus slow down further formation of oxide film [23], later reports showed that the corrosion resistance will decrease drastically by CNT addition since carbon act as cathode and accelerates the Mg dissolution [24-26]. The effect of CNT addition on the corrosion of Mg base composite is presented on Fig. 3. As it can be observed microgalvanic corrosion is increasing by increase in CNT content within the composite.

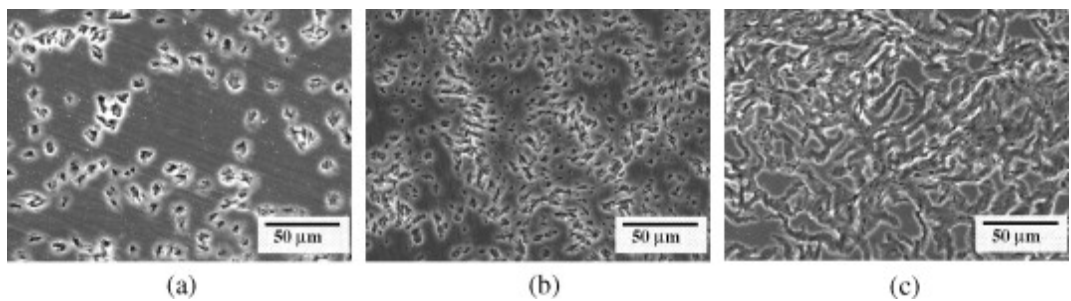

Figure 3. SEM micrographs showing corroded surface of Mg–CNT samples in 3.5 wt.% NaCl solution for 15 min: (a) Mg, (b) Mg–0.3CNT and (c) Mg–1.3CNT [24].

Fukuda et al. also observed that the corrosion initiates from the boundary of CNT and matrix (Fig. 4). Moreover CNT can inhibit the alloy from forming a protective oxide layer (passivation) [25]. Turhan et al. [26] reported that CNT addition will result in a thinner oxide layer since these fillers will promote hydrogen evolution which will detach loose oxides from the surface and hinders metal ions transport to surface.

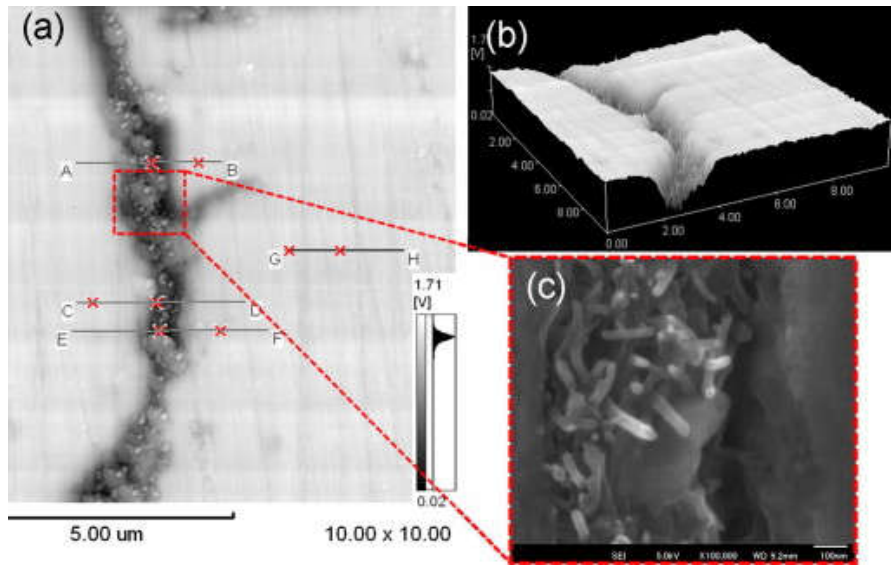

Figure 4. The 2D and 3D surface potential distribution of sample c, (a) and (b), respectively, and CNTs at the primary particle boundary, (c). Large SPD was observed between Mg matrix and CNTs [25].

There are a few works that has studied the effect of CNT on corrosion resistance of Zn base composites. Electrodepsoition is the most common processing method. Praveen et al. [27] studied the corrosion properties of Zn-CNT composite by polarization, gravimetric, salt spray and EIS tests and observed that acid treated CNTs enhances zinc corrosion resistance. SEM images electrodeposited samples and corrosion morphology of samples after 15 days in 3.5 wt% NaCl solution showed that the Zn-CNT coating has a more uniform and dense surface before and after corrosion (Fig. 5). Also ferroxyl test was used to assess the porosity of the coatings and it showed that using the CNT decreased the porosity of coatings. Therefore improvement in corrosion resistance was attributed to the role of CNT on obtaining a uniform coating with less defects and gaps. The same behavior was observed for Zn-Ni-CNT coating [28].

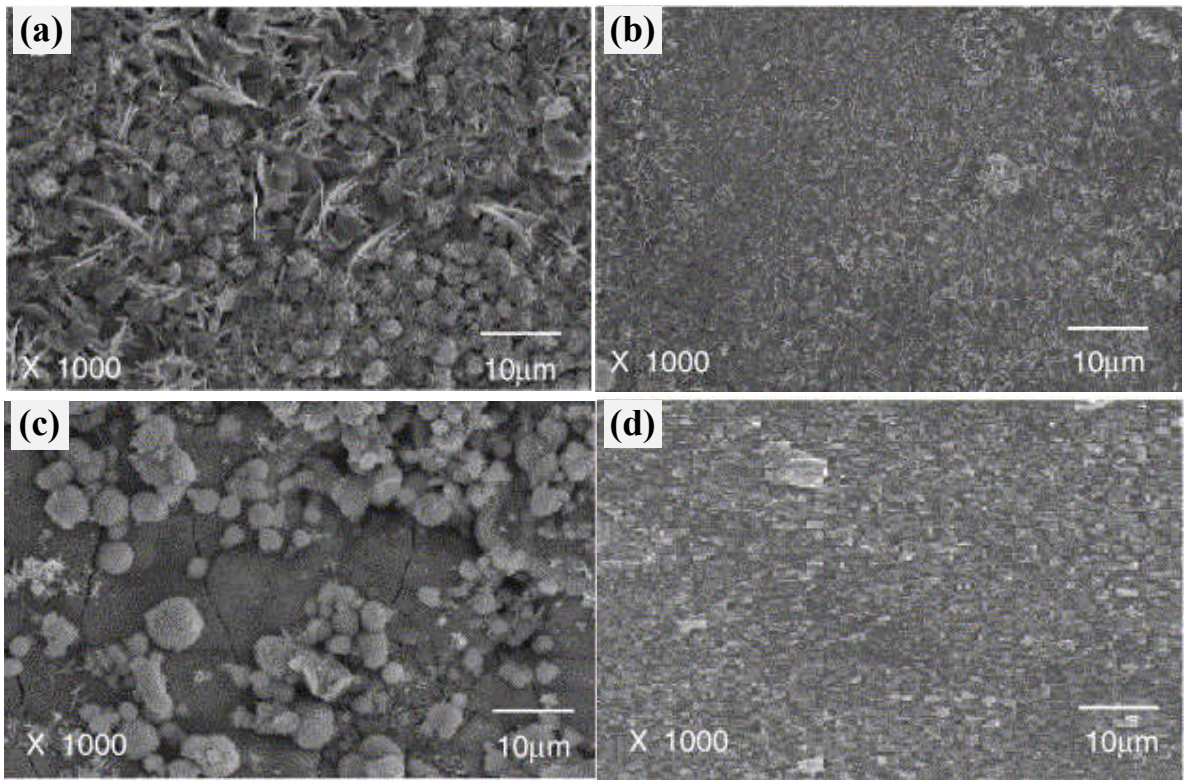

Figure 5. SEM images of electrodeposited (a) Zn, (b) Zn-CNT, and oxide layer surface after 15 days weight loss measurements of (c) Zn, (b) CNTs–Zn composite coatings [27].

In a recent study the corrosion properties of Cr-CNT produced by electrodeposition was assessed. It was observed that CNT addition broadened the passivation range. The reason provided is vague. Also uniform distribution of nanoparticles (CNT) results in uniform current distribution. Finally it was claimed that it is possible that some intermetallic compounds are formed [29].

Han et al. [30] tried to tackle the weak interaction between Sn solder alloy and CNT by coating the filler with Ni using electroless plating. Although they did not study the effect of applying coating on the surface of CNT, they reported that Ni coated CNTs improved the corrosion resistance. They reached to this conclusion by corrosion and pitting potentials shift to higher values which cannot be a proper criteria for claiming improvement in corrosion resistance. They also observed that the composite structure became smaller with increasing Ni-CNTs content which can result in more adherent and compact passivation layer. Furthermore, formation of a corrosion resistant passive layer of Ni-CNT and reducing galvanic corrosion between Sn (anode) and $Ag_3Sn$ (cathode) which is a common phenomenon in Sn-Ag-Cu solders in presence of Ni-CNT filler are other reasons brought by authors to back up their claim.

In a recent work, Dlugon et al. [31] fabricated a CNT coating on the surface of Ti by EPD and observed that the corrosion of Ti will increase by addition of CNT. No cause for such a behavior was proposed however, obtaining a dense and uniform CNT coating is very unlikely, therefore some areas of the Ti substrate are exposed to the corrosive solution which will promote crevice corrosion. A defective coating not only will not help the corrosion resistance but also make it more severe.

Generally the influence of CNT on corrosion of MMC includes two aspects: 1- Deterioration of corrosion resistance: CNTs has a higher standard potential compared with almost all the metals, therefore it will promote galvanic corrosion leading to acceleration in corrosion of surrounding metal. Also it will induce more boundaries which are more susceptible to corrosion. Especially in electrochemically deposited coatings, if the interfacial bonding is not strong enough, they will increase the roughness and defects in the matrix and therefore reduce the corrosion resistance. 2- Improvement of corrosion resistance: CNTs can fill the gaps and defects and act as physical barrier since they are chemically inert. Moreover, due to high standard potential, in metals with passivation capability they can promote formation of a protective oxide/passive layer. Finally, by acting as proper nucleation sites, the structure will be finer. It has been shown that in nanostructured coatings grain boundaries act as proper sites for nucleation and growth of passive layers. As a result grain refinement can result in uniform passivation and hence enhancement of corrosion resistance [32].

## 4. Effect of CNT on corrosion resistance of polymer matrix Composites

Applying organic coatings is one of the most common methods for protection of metals since they provide a physical barrier against corrosive species. These coatings are generally consist of three layers: primer, intermediate and top coat. The primer is conventionally made of zinc rich (and more recently Mg) epoxy composite. They key issue is forming a conductive network of Zn in epoxy which can connect the substrate to the particles and facilitate the cathodic protection. Therefore, zinc content is usually more than 80 wt%. Due to environmental regulations, and decrease in mechanical properties of the coating, there have been efforts to decrease the amount of toxic Zn from the coating. Graphite and carbon black has been used in these coatings to improve the conductivity [33, 34]. Compared to CNTs, these materials are chemically more active, have inferior mechanical and electrical properties and most importantly need higher percolation thresholds. Furthermore, epoxy has hydrophilic chemical groups such as hydroxyl groups (OH), carboxyl groups (C=O) which is not suitable for corrosion protection applications. On the other hand it has been shown that CNT can provide hydrophobic properties [35, 36]. Therefore using CNT in rich Zn or Mg primers has attracted the attention of researchers recently. Khun et al. [37] studied the effect of CNT content on corrosion resistance of epoxy coating on AA2024-T3. EIS results (Fig. 6) showed that by increasing the CNT content the coating the pore resistance of the epoxy will improve which shows ionic conductivity is impeded. Also CPE1 which indicates the permeation of the electrolyte is decreased by using CNT.

Finally R_ct which indicates the resistance to anodic dissolution of the substrate has increased. Hence it was deduced that CNT introduction improved the corrosion protective performance of the epoxy coating. This improvement was attributed to reduction of ionic conduction paths of epoxy by introduction of MWNTs. In other words CNTs reduced the through-porosity of the epoxy coating.

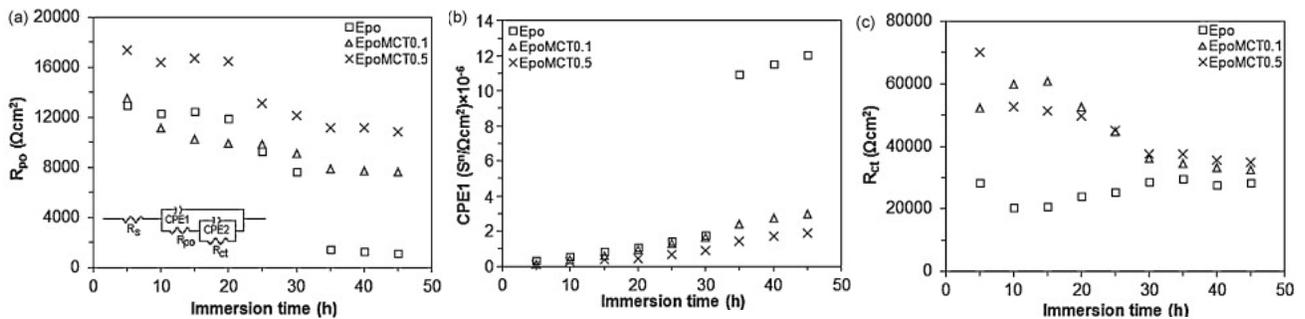

Figure 6. (a) Pore resistance (Rpo), (b) constant phase element (CPE1) and (c) charge transfer resistance (Rdl) measured in 0.5 M NaCl solution using EIS, as a function of immersion time. The inset in (a) show an equivalent circuit [33].

A few detailed work on corrosion protection of zinc rich epoxy coatings containing CNT have been published by Gergely group even though they do not study the effect of CNT directly [38, 39]. They observed that CNT addition to Zn 90 wt%-epoxy coating will decrease the charge transfer resistance gradually but if the Zn content is reduced to 70 wt% charge transfer resistance is very stable over time despite no improvement in galvanic protection ability was achieved. Generally they believe CNT will increase the corrosion rate of Zn particles and substrate since carbon has a more positive standard potential which leads to galvanic corrosion. In contrary, Park and Shon [40] reported that CNT introduction to Zn rich epoxy not only will enhance the conductivity and adhesion strength of ZRP but also will improve the corrosion resistance of the coating. As presented in Fig. 7, EIS results showed that total impedance at low frequencies which corresponds to corrosion resistance has increased with CNT and Zn content.

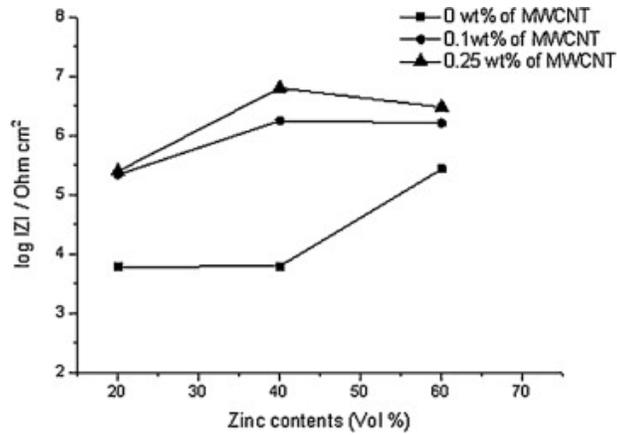

Figure 7. Impedance modulus of log |Z| at 0.01 Hz for epoxy zinc coated carbon steel without and with addition of MWCNTs [40].

Salt spray results were also in agreement with EIS results (Fig. 8) however no mechanism or reason for this behavior is suggested. Unfortunately the corrosion discussions and adhesion studies are very cursory.

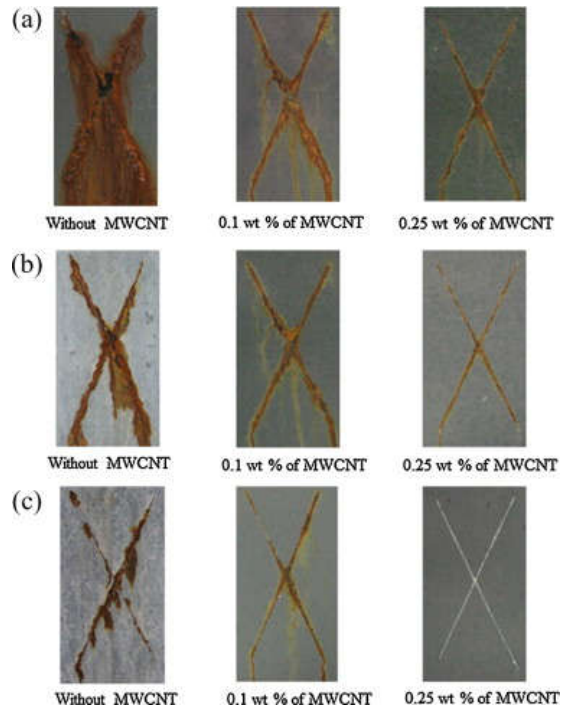

Figure 8. Corroded surface of epoxy zinc coated carbon steel with various contents of zinc and MWCNTs; (a) 20 vol% of zinc, (b) 40 vol% of zinc and (c) 60 vol% of zinc [40].

## 5. Conclusion

By development of new material, new promising coatings for corrosion protection are developed. With respect to the discussions and literature survey provided above, application of CNT in composites and coatings for corrosion protection is still at its initial stages. There are scarce papers, that has studied this field and most of them are very cursory. For metal-CNT category, Ni base composites have been studied relatively more. These research generally showed that CNT introduction improved the corrosion resistance by filling the gaps and coating defects, forming a passive layer of CNT, and promoting passive layer formation. The effect of CNT on the microstructure is not well studied. As mentioned before, for metals which can form a passive layer, finer grain size can lead to better corrosion resistance since formation of a dense passive layer is facilitated. On the other hand, for Mg base composites, CNT usually decreases the corrosion resistance. This is attributed to Carbon's high standard potential which will promote galvanic corrosion and to accelerate corrosion of surrounding metal. Also it will induce more boundaries which are more susceptible to corrosion and causes discontinuity in oxide layer. It has been shown that strength of interfacial bonding (which is a result of CNT functionalizing method) plays a major role on corrosion resistance of the composite.

For polymer base composites, application of CNT is inspired by the need to reduce metallic particles within the coating, CNT's high mechanical properties, and its very low conductivity threshold. However, there a just a few papers that has studied the effect of CNT on corrosion resistance of the coating and even less papers on ZRPs. Usually for polymer-CNT coating CNT will improve the corrosion resistance but the results for ZRP coatings contradicts each other. To evaluate the effect and realizing the mechanism it is necessary to do the EIS tests in long periods. Also salt spray tests and adhesion tests for organic coatings are also very crucial.